\documentclass[aps,amsmath,amssymb,superscriptaddress,twocolumn]{revtex4}
\usepackage{graphicx}

\begin{document}

\title{Is Hilbert space discrete?}

\author{Roman~V.~Buniy} \email{roman@uoregon.edu}
\affiliation{Institute of Theoretical Science \\ University of Oregon,
Eugene, OR 97403}

\author{Stephen~D.~H.~Hsu} \email{hsu@duende.uoregon.edu}
\affiliation{Institute of Theoretical Science \\ University of Oregon,
Eugene, OR 97403}

\author{A.~Zee}
\email{zee@itp.ucsb.edu}
\affiliation{Kavli Institute for Theoretical Physics \\
UCSB, Santa Barbara, CA 93106}

\begin{abstract}
We show that discretization of spacetime naturally suggests
discretization of Hilbert space itself. Specifically, in a universe
with a minimal length (for example, due to quantum gravity), no
experiment can exclude the possibility that Hilbert space is
discrete. We give some simple examples involving qubits and the
Schrodinger wavefunction, and discuss implications for quantum
information and quantum gravity.
\end{abstract}

%\pacs{}

\maketitle

\date{today}

There are many indications that spacetime may be discrete rather than
continuous \cite{minlength}.  For example, metric fluctuations due to
quantum gravity might preclude any notion of distances less than of
order the Planck length $l_P$.  Recent work \cite{CGH} has shown that
no macroscopic experiment can be sensitive to discreteness of position
on scales less than the Planck length. Any device (such as an
interferometer) capable of such resolution would be so massive that it
would have already collapsed into a black hole. Relativistic
covariance suggests that discretization of space implies
discretization of time. Indeed, minimal length probably makes it
impossible to construct a clock capable of measuring time differences
less than of order the Planck time. (Consider, for example, a bouncing
photon between two mirrors as the ticking of the clock.) As an
explicit but crude toy model of discreteness, one might imagine that
our universe lives on a spacetime lattice with spacing $l_P$. More sophisticated models have been proposed \cite{discretemodels} in
which spacetime is discrete, but not necessarily regular. In
our discussion discreteness should not be taken to imply regularity,
either in spacetime or the structure of Hilbert space. We are not
suggesting that continuous Hilbert space necessarily be replaced by a
lattice; instead, for example, the discreteness might be due to an
intrinsic fuzziness or uncertainty.

A consequence of spatial discreteness is that in any finite region of
space of size $L$ there are only a finite number of degrees of freedom
$N \sim L^3$.  (Henceforth we adopt units in which $l_P = 1$.)
Although our universe might be infinite in extent, any experiment
performed by scientists must take place over a finite period of
time. By causality, this implies that the experiment takes place in a
region of finite size, which we take to be $L$. We therefore assume
the existence of a long-distance (infrared) regulator $L$ in addition
to a short-distance (ultraviolet) regulator $l_P$.

Now consider quantum mechanics in a spatially discrete universe. The
dimensionality of Hilbert space is itself finite, equal to the number
of degrees of freedom. Let the space be spanned by a finite set of
independent basis vectors $\vert n \rangle$ ($1 \leq n \leq N$).  In
conventional quantum mechanics, we define Hilbert space to consist of
all linear combinations of these basis vectors
\begin{equation}
\label{psi}
\vert \psi \rangle = \sum^N_{n=1} a_n \vert n \rangle ,
\end{equation}
modulo rescaling by an arbitrary complex parameter. Since the $a_n$
are continuous complex parameters, Hilbert space is continuous even if
spacetime is discrete, and the set of possible states $\vert \psi
\rangle$ is infinite.

However, in a spatially discrete universe there is no experiment which
can exclude discreteness of the coefficients $a_n$, if that
discreteness is sufficiently small. We argue as follows. If the number
of degrees of freedom is finite, so is the set of possible distinct
measurement devices one can construct. (By ``distinct'' devices we do
not mean different in design or construction, but rather that they
measure distinct physical quantities -- in other words, correspond to
different operators acting on the Hilbert space. See the qubit example
below.) Equivalently, the number of eigenstates of all possible
distinguishable operators is finite (recall that with ultraviolet
and infrared regulators present, the spectrum, and hence the
number of eigenstates, of any particular operator is finite). Thus,
the physics of this universe can be described using a Hilbert space
with only a finite number of distinguishable states -- that is, a
discrete and finite Hilbert space, in which the values of $a_n$ are
themselves quantized.

As a simple example, consider a single qubit. The Hilbert space of a
spin-$\frac{1}{2}$ particle is simply the set of all eigenstates of
the spin operator. The most general state $\vert \psi \rangle$ can be
written as
\begin{equation}
\label{qubit}
\vert \psi \rangle = \cos \theta \, \vert + \rangle \, + \, 
e^{i\phi} \sin \theta \, \vert - \rangle,
\end{equation}
where $\theta$ and $\phi$ are continuous parameters. However, discrete
space implies that there are only a finite number of distinguishable
spin operators. A sufficiently small rotation of the measurement
apparatus is indistinguishable from no rotation. Hence, one cannot
measure changes in the angular variables smaller than $\epsilon \sim 1 / d$, where $d$ is the size of the apparatus. This size is
somewhat ill-defined, since by making the apparatus arbitrarily long
it becomes sensitive to very small rotations (neglecting, of course, considerations of rigidity and causality). One should probably take $d$ to describe only the part of the
apparatus with which the qubit interacts during a measurement.
Alternatively, if the spin-$\frac{1}{2}$ object described by $\vert
\psi \rangle$ has finite size (Compton wavelength), the set of its
possible orientations might itself be discrete (imagine a vector
constrained to connect two vertices of a lattice). In that case $d$
might be given by the size of the qubit, rather than that of the
apparatus.

\begin{figure}[h]
\includegraphics[width=4cm]{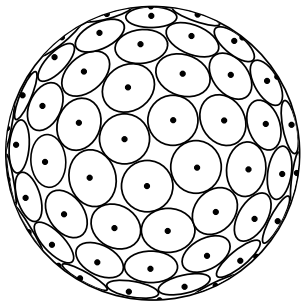}
\caption{A possible discretization of the Bloch sphere (qubit Hilbert
  space). Points on each disc (of size $\epsilon$) are
  identified. Points between discs can be assigned to the nearest disc.
}
\label{figure}
\end{figure}

Discrete Hilbert space leads us to a concrete modification of the
linear superposition principle. For example, if we were to superpose
two states of the form (\ref{qubit}), one with $( \theta_1, \phi_1)$
and the other with $( \theta_2, \phi_2)$, then for arbitrary choice of
coefficients the resulting state $(\theta, \phi)$ will not be in the
allowed set. One concrete proposal would be to replace $(\theta,
\phi)$ by the nearest allowed values (a ``snap to nearest lattice
site'' rule; see Fig.~\ref{figure}). We imagine that a clever
experimentalist could set a useful bound on this deviation from linear
superposition.

One might be concerned that the SU(2) group structure of qubit
rotations cannot be obtained as the limit of larger and larger finite
discrete subgroups. However, there exist simple models in which
continuous rotational or even Lorentz symmetry is obtained in the long
wavelength limit from underlying dynamics which has only discrete
symmetry. For example, in lattice QCD the symmetries are all discrete,
yet continuous symmetries emerge in the long wavelength limit. As
another example, in \cite{zee} a model of spinless point particles
hopping on a flux lattice gives rise to low-energy excitations obeying
the Dirac equation.

We can deduce similar results concerning discreteness by considering
the Schrodinger equation. All the physics of a universe with discrete
spacetime can be described by a wavefunction $\psi (x)$ whose values
are discrete, rather than continuous. To convince ourselves of this,
we need merely consider simulations of the Schrodinger equation on a
classical digital computer, in which both the spacetime coordinates
and $\psi(x)$ are discrete. All predictions of quantum mechanics can
be obtained to any desired accuracy using such discrete simulations;
quantum phenomena such as interference patterns can therefore be
reproduced even if quantum mechanics is intrinsically discrete, as
suggested here. Since by assumption we can probe the variation of
$\psi(x)$ only over distances larger than the Planck length, the
magnitude of the required accuracy $\Delta \psi$ is bounded
below. (Again, since we have both ultraviolet and infrared regulators,
$\psi(x)$ must be a smooth function. It cannot vanish identically in
an entire region, so its variation over some finite interval is
bounded below.)  It is important to note, though, that the required
discreteness might be exponentially small. For example, to describe
the exponential tail of a wavefunction might require $\Delta \psi \sim
e^{- L }$, where $L$ is the box size of the simulation. Nevertheless,
for fixed $L$ and $l_P$, the magnitude of $\Delta \psi$ is always
bounded below.

A discrete wavefunction $\psi(x)$ {\it implies} a discrete Hilbert
space, and vice versa, since the value of the wavefunction is simply
the overlap of a particular state, $\vert \psi \rangle$, with another,
$\vert x \rangle$. In other words, using Eq.~(\ref{psi}), $\psi(x_i) =
\langle x_i \vert \psi \rangle = a_i$.  If the set of states $\vert
\psi \rangle$ and the set of states $\vert x \rangle$ are both finite,
then $\psi (x)$ can take only a finite (discrete) set of values, and
vice versa. As discussed above, to accommodate exponential fall-off
the size of discreteness in the $a_i$ must be exponentially
small---potentially of order $e^{-L}$, where $L$ is the size of the
universe.

However, minimal length seems to imply a stronger (non-exponential)
limitation on the phase information carried by a quantum state,
similar to what we obtained above for a qubit. Suppose information is
stored in a particular quantum state $\vert \psi \rangle$. A
Planck-length uncertainty in the spacetime location of the state (or
of where the measurement of the state takes place) leads to an
uncertainty in the value of the phase, as seen from the time
translation operator $U(t) = e^{-iHt}$ or the translation operator
$T(x) = e^{-ipx}$. The phase can be specified only to accuracy $E_*$
or $p_*$ in Planck units, where $E_*$ and $p_*$ are roughly the
characteristic energy or momentum associated with $\vert \psi
\rangle$. To be explicit, one can expand $\vert \psi \rangle$ in an
energy or momentum eigenstate basis, with each term in the expansion
acquiring a phase uncertainty of order $E$ or $p$.  Using $p > 1/L$,
we obtain a phase uncertainty $\epsilon > 1 / L$, similar to $1 / d$
in the case of a spin-$\frac{1}{2}$ qubit. If the state $\vert \psi
\rangle$ is transported over some path in spacetime, for example to be
interfered with some other state, we expect a fundamental limitation
on the precision of the relative phase. This might have some
interesting consequences for Berry's phase.

We can derive this result another way by considering a particle of
energy $E$, interacting with an external probe of energy $E'$ which
measures the phase $\phi$ of its wavefunction. Let the interaction
take place over a time interval $\Delta t$. The phase of the particle
necessarily evolves during the time interval, so $\Delta \phi \sim E
\Delta t$. Causality requires $\Delta t > R$, where $R$ is the size of
the probe (or the portion of it which interacts with the particle,
which we take to be the entire probe). It must be the case that $R >
E'$, or the probe would have already collapsed into a black
hole. Finally, using energy-time uncertainty, $E' > (\Delta t)^{-1}$ ,
we obtain
\begin{equation}
\Delta \phi \sim E \Delta t > E E' > E^2 / \Delta \phi,
\end{equation}
which implies $\Delta \phi > E$. So, we expect that a phase
discreteness $\epsilon$ smaller than $\sim E$ (recall, we use Planck
units) is undetectable experimentally. 

The phase uncertainty discussed above is not
inconsistent with the requirement that $\Delta \psi$ might be
exponentially small. There is no contradiction between an
exponentially small uncertainty in the magnitude and larger
uncertainty in the phase of a wavefunction $\psi$. For example,
suppose $\psi$ is expressed as the superposition of two other states:
$\psi = a_1 \psi_1 + a_2 \psi_2$. Our ability to measure $\vert \psi
\vert$ (or $\vert \psi_i \vert$) to arbitrary accuracy places no
limitation on the precision of the phases in $a_i$ or $\psi_i$.

We mention some consequences of discrete Hilbert space:
\begin{enumerate}
\item Only a finite number of classical bits are required to specify
the state of a discrete qubit.  Note that, because we cannot directly measure
its state, a single qubit can be used to transmit or store only a
single bit of classical information. (This is a result of Holevo's
theorem \cite{holevo}.) Nevertheless, a perfect classical simulation of
qubits with continuous Hilbert space requires an infinite number of
classical bits.

\item There are asymptotic limits to the power of quantum
computation. Consider, for example, Shor's algorithm \cite{Shor} (or
similarly the quantum Fourier transform), which requires of order
$(\ln n)^3$ steps to factor an integer $n$. Discreteness of order
$\epsilon$ limits the precision of quantum manipulations, and quantum
algorithms require a minimum precision of order one over the number of
steps in the algorithm \cite{BBBV}.  Thus, factorization of an integer
$n$ using Shor's algorithm requires manipulations of precision $(\ln
n)^{-3}$, and there exists a largest integer $n_* \sim \exp (
{\epsilon}^{-\frac{1}{3}})$ that can be factored. In the case of
arbitrarily low-energy qubits, we might have $\epsilon \sim 1/L \sim
1/T$, where $T$ is the timescale of the quantum processor, so $n_*$
still grows exponentially with $T^{\frac{1}{3}}$. However, for qubits
of fixed energy $E$
%(affected, for example, by the ambient temperature), 
one eventually encounters a maximum $n_*$.

\item Quantum mechanics is modified at short distances. It may be that
near the Planck scale the Hilbert space
discreteness $\epsilon$ is of order unity. For example, a vector whose
length $d$ is of order the Planck length may have only a few possible
orientations (imagine that the vector must connect two points on a
lattice). According to this analogy, quantum dynamics might be
drastically modified at short distances. It would be interesting to
formulate a superclass of models of this type which have ordinary
quantum mechanics as a limiting case. These might produce a novel
approach to quantum gravity, as current approaches such as string
theory extrapolate quantum mechanics with continuous Hilbert space all
the way to the Planck scale.
\end{enumerate}

\bigskip

One might ask how to evolve a state in a discrete Hilbert space.
There are many possibilities, but one concrete method would
be to write the time evolution operator $e^{-iHt}$ as a product
of discrete evolution operators $e^{-i H \Delta t}$ 
and apply this product of operators sequentially
to the state, for example as in (2), followed by the ``snap to''
rule after each step.  This is equivalent to taking classical digital
computer simulations literally. That is, by accepting the finite
precision of the variable $\psi (x)$ in an ordinary computer program,
one obtains a naive discretization of Hilbert space with the ``snap
to'' rule implemented by simple numerical rounding.

In conclusion, it appears that the traditional assumption of
continuous Hilbert space is rather strong: minimal length precludes
any experiment showing that the discreteness parameter $\epsilon$ is
exactly zero. While we have motivated a non-zero $\epsilon$ using
quantum gravity, we stress that discreteness may appear at a dimensional
scale larger than $l_P$, and that experimentalists should keep an open mind.

\bigskip
\emph{Acknowledgements.---} The authors thank D.~Bacon, M.~Graesser,
B.~Murray and J.~Polchinski for useful comments. S.~H. and R.~B. are
supported by the Department of Energy under
DE-FG06-85ER40224. A.~Z. was supported in part by the National Science
Foundation under grant number PHY 99-07949(2004).

%%%%%%%%%%%%%%%%%%%%%%%%%%%%%%%%%%%%%%%%%%%%%%%%%%%%%%%%%%%%%%%%%
%%%
%%%                     BIBLIOGRAPHY
%%%
%%%%%%%%%%%%%%%%%%%%%%%%%%%%%%%%%%%%%%%%%%%%%%%%%%%%%%%%%%%%%%%%%

\bigskip

%\newpage
%\vskip .75 in

\end{document}